\documentclass[]{aa}  

\usepackage{graphicx}
\usepackage{color}
\usepackage{txfonts}

\begin{document}

   \title{The ALMA-PILS survey: first detection of methyl isocyanide (CH$_3$NC) in a solar-type protostar}
   \titlerunning{High-resolution studies of complex cyanides}

   \author{H. Calcutt\inst{1},  M. R. Fiechter\inst{1,2}, E. R. Willis\inst{3}, H. S. P. M\"uller\inst{4}, R. T. Garrod\inst{3}, J. K. J{\o}rgensen\inst{1}, S. F. Wampfler\inst{5}, T. L. Bourke\inst{6}, A. Coutens\inst{7}, M. N. Drozdovskaya\inst{5}, N. F. W. Ligterink\inst{8,9}, L. E. Kristensen\inst{1}}

\institute {\inst{1}Centre for Star and Planet Formation, Niels Bohr Institute \& Natural History Museum of Denmark, University of Copenhagen,\\ \phantom{'}{\O}ster Voldgade 5--7, DK-1350 Copenhagen K., Denmark, \email{calcutt@nbi.ku.dk}\\
\inst{2}University of Groningen, Nijenborgh 4, 9747 AG, Groningen, The Netherlands\\    
\inst{3}Departments of Chemistry and Astronomy, University of Virginia, Charlottesville, VA 22904, USA\\
\inst{4}I. Physikalisches Institut, Universit\"at zu K\"oln, Z\"ulpicher Str. 77, 50937 K\"oln, Germany\\
\inst{5}Center for Space and Habitability, University of Bern, Gesellschaftsstrasse 6, CH-3012 Bern, Switzerland\\
\inst{6}SKA Organization, Jodrell Bank Observatory, Lower Withington, Macclesfield, Cheshire SK11 9DL, UK\\
\inst{7}Laboratoire d'Astrophysique de Bordeaux, Univ. Bordeaux, CNRS, B18N, all\'{e}e Geoffroy Saint-Hilaire, 33615 Pessac, France\\     
 \inst{8}Raymond and Beverly Sackler Laboratory for Astrophysics, Leiden Observatory, Leiden University, PO Box 9513, 2300 RA Leiden,\\ \phantom{'}The Netherlands\\      
\inst{9}Leiden Observatory, Leiden University, PO Box 9513, 2300 RA Leiden, The Netherlands\\
}
\authorrunning{Calcutt et al.}

   \date{Received}
 
  \abstract
   {Methyl isocyanide (CH$_3$NC) is the isocyanide with the largest number of atoms confirmed in the interstellar medium (ISM), but it is not an abundant molecule, having only been detected towards a handful of objects. Conversely, its isomer, methyl cyanide (CH$_3$CN), is one of the most abundant complex organic molecules detected in the ISM, with detections in a variety of low- and high-mass sources.}
   { The aims of this work are to determine the abundances of methyl isocyanide in the solar-type protostellar binary IRAS 16293--2422 and understand the stark abundance differences observed between methyl isocyanide and methyl cyanide in the ISM. }
   { We use ALMA observations from the Protostellar Interferometric Line Survey (PILS) to search for methyl isocyanide and compare its abundance with that of its isomer methyl cyanide. We use a new line catalogue from the Cologne Database for Molecular Spectroscopy (CDMS) to identify methyl isocyanide lines. We also model the chemistry with an updated version of the three-phase chemical kinetics model {\em MAGICKAL}, presenting the first chemical modelling of methyl isocyanide to date. }
   {We detect methyl isocyanide for the first time in a solar-type protostar, IRAS 16293--2422 B, and present upper limits for its companion protostar, IRAS 16293--2422 A. Methyl isocyanide is found to be at least 20 times more abundant in source B compared to source A, with a CH$_3$CN/CH$_3$NC abundance ratio of 200 in IRAS 16293--2422 B and >5517 in IRAS 16293--2422 A. We also present the results of a chemical model of methyl isocyanide chemistry in both sources, and discuss the implications on methyl isocyanide formation mechanisms and the relative evolutionary stages of both sources. The chemical modelling is unable to match the observed CH$_3$CN/CH$_3$NC abundance ratio towards the B source at densities representative of that source. The modelling, however, is consistent with the upper limits for the A source. There are many uncertainties in the formation and destruction pathways of methyl isocyanide, and it is therefore not surprising that the initial modeling attempts do not reproduce observations. In particular, it is clear that some destruction mechanism of methyl isocyanide which does not destroy methyl cyanide is needed. Furthermore, these initial model results suggest that the final density plays a key role in setting the abundance ratio. The next steps are therefore to obtain further detections of methyl isocyanide in more objects, as well as undertaking more detailed physico-chemical modeling of sources such as IRAS16293.}
   {}


   \maketitle
%
\section{Introduction}
Methyl isocyanide (CH$_3$NC) is the isocyanide with the largest number of atoms confirmed in the interstellar medium (ISM), however, its detections are far and few between. It was first tentatively detected by \citet{Cernicharo1988} in the Sgr B2 cloud, and confirmed with additional transitions by \citet{Remijan2005}. Since then, it has only been confirmed in a handful of sources, including the Horsehead photo-dissociation region (PDR) by \citet{Gratier2013}, and Orion KL by \cite{Lopez2014}. Conversely, its isomer methyl cyanide (CH$_3$CN) is one of the most abundant complex organic molecules detected in the ISM, with detections in a variety of low- and high-mass sources (e.g. \citealt{Bisschop2008, Beltran2005, Zapata2010}). 

Such stark differences in the abundances of two isomers raises interesting questions about their chemistry in the ISM, and in particular, whether their abundance differences are a consequence of their formation pathways, a result of differential destruction with chemical evolution, or even other processes. A number of authors have explored this with both theoretical and laboratory work. Theoretical calculations by \citet{Defrees1985} of the CH$_3$NC/CH$_3$CN ratio in dense interstellar clouds explored the ratio which is inherited from their protonated precursor ions, protonated methyl cyanide (CH$_3$CNH$^+$) and protonated methyl isocyanide (CH$_3$NCH$^+$). These ions are formed when CH$_3$$^+$ reacts with HCN, forming a complex which can rapidly equilibrate between the two isomeric forms. As the complex relaxes to form a stable molecule, isomerisation occurs until the internal energy of the ion is no longer sufficient to overcome the isomerisation barrier. Ion-electron recombination reactions then lead to the formation of methyl isocyanide and methyl cyanide with methyl cyanide being formed in higher amounts than methyl isocyanide. Methyl cyanide is preferred because it is the more stable of the two isomers. 

Laboratory experiments by \citet{Hudson2004} looked at the effect of non-thermal processes on the formation of isocyanides. They found that bombarding pure ice samples of methyl cyanide with protons and UV photons produced methyl isocyanide, however, the introduction of water-rich ices inhibits the process, with OCN$^-$ being formed instead. \citet{Mencos2016} also used laboratory experiments to show that methyl isocyanide could be formed through the reaction between methyl cyanide and N atoms in ice, triggering isomerisation that produces methyl isocyanide and ketenimine (CH$_2$CNH) as principal products, with methyl isocyanide being formed in higher amounts.

As well as formation and destruction mechanisms, differences in adsorption properties could be crucial to understanding the observed differences in abundance between methyl cyanide and its isocyanide. \citet{Bertin2017a, Bertin2017b} used temperature programmed desorption (TPD) experiments to study the adsorption of methyl cyanide and methyl isocyanide on interstellar grain surfaces. They found that for amorphous water ice a small amount of the adsorbed molecules desorb through a volcano effect, meaning that some diffusion takes place during the warming up and some of the methyl cyanide and methyl isocyanide molecules get trapped in the corrugation of the amorphous water bulk. The adsorption energies of methyl cyanide were found to be slightly higher than those of methyl isocyanide by 20 to 40 meV, both experimentally and theoretically. They attribute this to the slightly stronger ability of methyl cyanide to form H-bonds because of a local excess of positive charge on the hydrogen atoms of the methyl group. On graphite surfaces the adsorption energy of each molecule was found to be very sensitive to the structural defects of aromatic carbonaceous surfaces, increasing the average adsorption energy by more than 50\% compared to a perfect graphene plane. They again found that the adsorption energy was higher for methyl cyanide than its isomer, which they predict would lead to a gas-phase enrichment of the isonitrile at a given grain temperature compared with the abundance ratio of the two isomers in the condensed phase.

What is lacking in the study of methyl isocyanide in the ISM is its detection in more objects, to provide statistically significant abundance ratios and chemical modelling of its formation and destruction pathways in different environments. To achieve the first point, high sensitivity observations are needed as it has shown very weak emission in the objects it has been detected in so far. In this work, we therefore utilise high sensitivity ALMA observations from the Protostellar Interferometric Line Survey (PILS) to present the first detection of methyl isocyanide in a solar-type protostar, IRAS 16293--2422 (hereafter IRAS 16293). IRAS 16293 is a Class 0 solar-type protostellar binary, located only 141\,pc away \citep{Dzib2018}, in the Ophiuchi star-forming complex. It consists of at least two sources, A and B which are separated by 5\arcsec (705\,AU), and each show a rich spectrum of complex organic molecules \citep{Bisschop2008, Jorgensen2016, Lykke2017}.\\

We present details of the observations in Section \ref{sec:obs}, the results of an abundance analysis and chemical modelling in Section \ref{sec:res} and a discussion of the astronomical implications in Section \ref{sec:dis}.  Finally, the conclusions of this work are presented in Section \ref{sec:con}.

\section{Observations}\label{sec:obs}
The observations presented here were taken as part of the PILS program, an ALMA observing program to study IRAS 16293 in Band 7, between 329.147\,GHz and 362.896\,GHz. The phase centre of the observations is located between the two components of the binary system at $\alpha_{J2000}$ = 16$^{\rm h}$32$^{\rm m}$22{\rm \fs}72, $\delta_{J2000}$=$-$24$^{\circ}$28\arcmin34\farcs3. The data are a combination of the main 12\,m array and Atacama Compact Array (ACA) observations, which have been combined to have a restoring beam of 0\farcs5 at a spectral resolution of 0.2\,km\,s$^{-1}$. They reach a sensitivity of about 7--10\,mJy\,beam$^{-1}$\,channel$^{-1}$, i.e., approximately $\sim$5\,mJy\,beam$^{-1}$\,km\,s$^{-1}$ across the entire frequency range. Further details of the data reduction and continuum subtraction procedure can be found in \citet{Jorgensen2016}. \\

In this work we analyse spectra extracted from two positions, where a large number of complex organic molecules have been previously detected and the line emission is not obscured by dust. The first position at $\alpha_{J2000}$=16$^{\rm h}$32$^{\rm m}$22{\rm \fs}58, $\delta_{J2000}$=$-$24$^{\circ}$28\arcmin32\farcs8, is offset from the continuum peak position of IRAS 16293B in the south west direction by one beam diameter (0\farcs5, 70\,AU). This position is used as the lines are particularly bright, do not have strong absorption features, and do not suffer from high continuum optical depth (\citealt{Coutens2016}, \citealt{Lykke2017}). The FWHM of lines around source A are much broader ($\sim$2$-$3\,km\,s$^{-1}$) than around source B ($\sim$1\,km\,s$^{-1}$) making line blending a significant problem for identification and analysis of molecular emission. We therefore choose a position at $\alpha_{J2000}$=16$^{\rm h}$32$^{\rm m}$22{\rm \fs}90, $\delta_{J2000}$=$-$24$^{\circ}$28\arcmin36\farcs2, that is 0\farcs6 (85\,AU) north east from the peak continuum position of IRAS 16293A. This position has been used previously to study the nitrile chemistry in IRAS 16293 \citep{Calcutt2017}, the oxygen-bearing molecules \citep{Manigand2018}, and methyl isocyanate emission \citep{Ligterink2017}. \\

\section{Results and analysis}\label{sec:res}

Both offset positions towards A and B were searched for methyl isocyanide. Isotopologues were not searched for as the line lists are not available. Methyl isocyanide is detected towards IRAS 16293B, with 10 unblended lines and 6 blended lines above 3$\sigma$ present in the spectra. All of the lines that are predicted by a local thermodynamic equilibrium (LTE) spectral model are detected in the spectra. A list of the detected transitions, frequencies, upper energy levels, Einstein coefficients and whether the transitions are blended are given in Table \ref{tab:spec}. This is the first detection of this molecule in a solar-type protostar. No lines are detected towards the offset position of IRAS 16293A. We additionally searched positions closer to the A peak continuum position, however, no lines could be identified as blending became too severe. \\

\begin{table}
\caption{Spectroscopic information for the lines of CH$_3$NC detected in IRAS 16293B.}\label{tab:spec}
\centering
\begin{tabular}{ccccc}
\hline
\hline
Transition&Frequency &$E\rm{_{u}}$ &$A\rm{_{ij}}$&Blended$^{\dagger}$\\
&(MHz)&(K)&(s$^{-1}$)&\\

\hline 
	
 17 7\,--\,16 7 &		341\,329.6&	493&	2.83$\times$10$^{-3}$&U\\	
 17 6\,--\,16 6 &		341\,429.4&	402&	2.98$\times$10$^{-3}$&B\\	
 17 5\,--\,16 5 &		341\,513.8&	324&	3.11$\times$10$^{-3}$&B\\	
 17 4\,--\,16 4 &		341\,582.9&	261&	3.22$\times$10$^{-3}$&U\\	
 17 3\,--\,16 3 &		341\,636.7&	211&	3.31$\times$10$^{-3}$&B\\	
 17 2\,--\,16 2 &		341\,675.1&	176&	3.37$\times$10$^{-3}$&B\\	
 17 1\,--\,16 1 &	       341\,698.2&	155&	3.40$\times$10$^{-3}$&U\\	
 17 0\,--\,16 0 &		341\,705.9&	148&	3.41$\times$10$^{-3}$&U\\	
 18 7\,--\,17 7 &		361\,396.3&	511&	3.43$\times$10$^{-3}$&U\\	
 18 6\,--\,17 6 &		361\,501.8&	419&	3.60$\times$10$^{-3}$&B\\	
 18 5\,--\,17 5 &		361\,591.2&	341&	3.74$\times$10$^{-3}$&U\\	
 18 4\,--\,17 4 &		361\,664.3&	278&	3.85$\times$10$^{-3}$&B\\	
 18 3\,--\,17 3 &		361\,721.2&	229&	3.94$\times$10$^{-3}$&U\\	
 18 2\,--\,17 2 &		361\,761.9&	193&	4.01$\times$10$^{-3}$&U\\	
 18 1\,--\,17 1 &		361\,786.3&	172&	4.04$\times$10$^{-3}$&U\\	
 18 0\,--\,17 0 &		361\,794.4&	165&	4.06$\times$10$^{-3}$&U\\		
	
\hline
		
	\end{tabular}
\tablefoot{$^{\dagger}$U denotes an unblended line and B denotes a blended line.}
\end{table}
\subsection{Spatial extent}

\begin{figure*} 
\begin{center} 
\includegraphics[width=15.2cm, angle=0, clip =true, trim = 1cm 0cm 1cm 0cm]{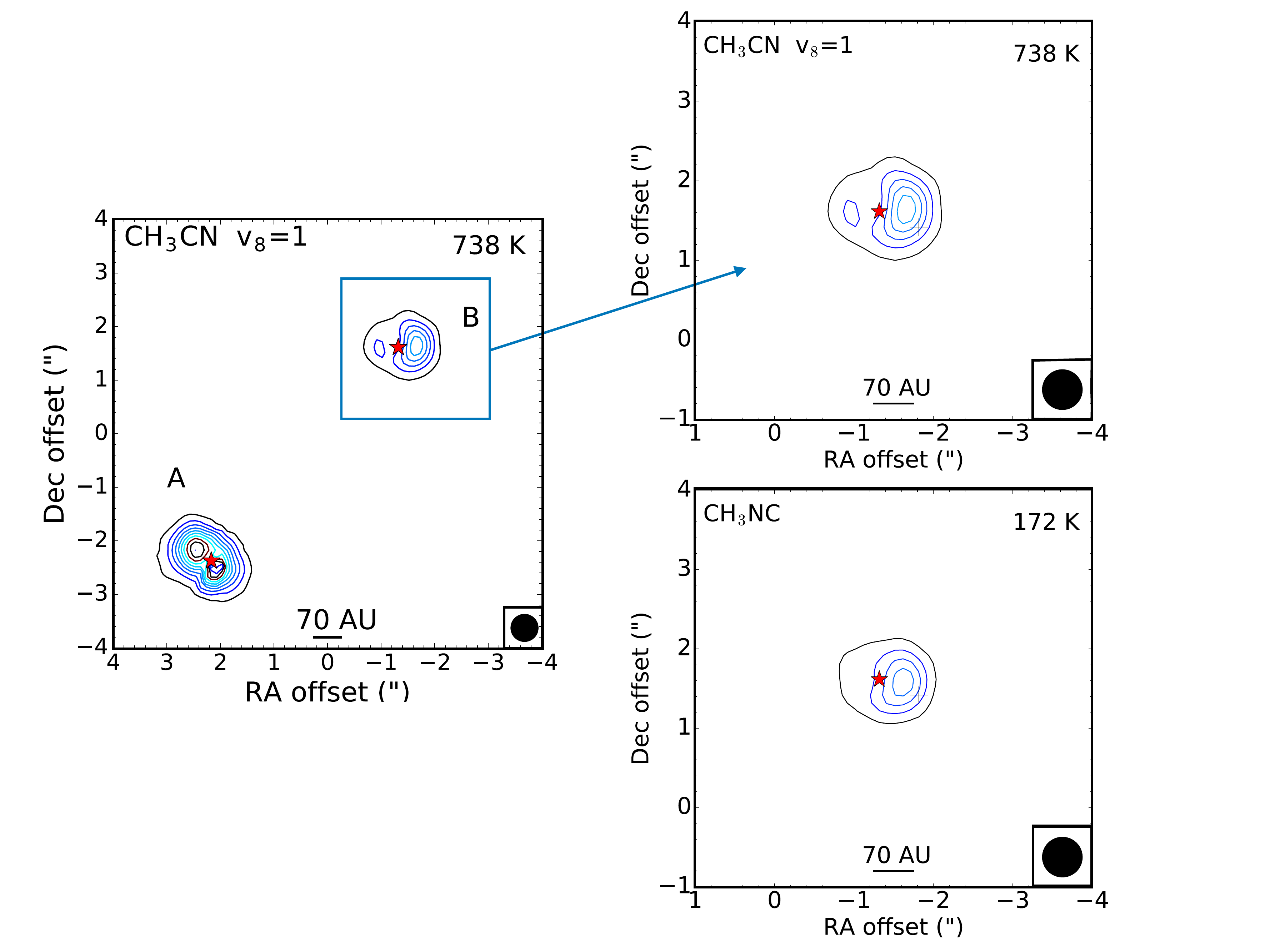}
\end{center} 
\caption{The panel on the left shows a velocity-corrected integrated emission (VINE) map of the $18_{\pm4} - 17_{\mp4}$ $\varv_8 = 1$ $I = +1$ transitions (332.029\,GHz) of methyl cyanide, which are unresolved. The right panel shows a zoom in of the methyl cyanide emission towards IRAS 16293B and an integrated emission map of the 18$_1$\,--\,17$_1$ $\varv$=0 line (361.786\,GHz) of methyl isocyanide. For IRAS 16293B six channels (1.2\,km\,s$^{-1}$) centered on the $V_{\rm LSR}$ are integrated over. For IRAS 16293A 16 channels (3.2\,km\,s$^{-1}$) centered on the $V_{\rm peak}$ of each pixel are integrated over. The axes show the position offset from the phase centre of the observations. Contour levels start at 30 mJy\,km\,s$^{-1}$ and increase in steps of 50 mJy\,km\,s$^{-1}$. The red star marks the peak continuum position in each source and the black cross marks the offset position where the spectra analysed in this work are extracted from. The RA and Dec offsets are relative to the phase centre of the observations.\label{fig:ch3ncmap}} 
\end{figure*}

Figure \ref{fig:ch3ncmap} shows emission maps for methyl cyanide and methyl isocyanide in IRAS 16293. In the left panel, a velocity-corrected integrated emission (VINE) map shows the emission from methyl cyanide (the $18_{\pm4} - 17_{\mp4}$ $\varv_8 = 1$ $I = +1$ transitions, which are unresolved). VINE maps are a method for determining the spatial scale of molecules in spectrally dense regions, with large velocity gradients. They remove the problem of contamination by nearby lines in the emission map. More details of how a VINE map is created and the caveats with this method can be found in \citet{Calcutt2017}. The right panel of Fig. \ref{fig:ch3ncmap} shows a zoom in of the methyl cyanide emission and an integrated emission map of methyl isocyanide (the 18$_1$\,--\,17$_1$ $\varv$=0 line). The emission traces the disk component of IRAS 16293B and shows a similar morphology and extent to methyl cyanide emission. The other lines of methyl isocyanide also show the same extent and morphology. \\

\subsection{Column densities and excitation temperatures}\label{sec:col}
To determine abundances and excitation temperatures, the spectral modelling software CASSIS\footnote{CASSIS has been developed by IRAP-UPS/CNRS: \url{http://cassis.irap.omp.eu/}} was used to calculate synthetic spectra and determine best-fit spectral models to the observations assuming LTE. A large grid of models was run with column densities between 1$\times$10$^{13}$\,--\,9$\times$10$^{14}$\,cm$^{-2}$ and excitation temperatures between 80\,--\,300 K. Line widths were varied between 0.8\,--\,1.2\,km\,s$^{-1}$, which are the typical line widths found in other PILS studies (e.g. \citealt{Jorgensen2016, Ligterink2017}). The peak velocities were varied between 2.5\,--\,2.9\,km\,s$^{-1}$, corresponding to one channel on either side of the $V\rm{_{LSR}}$ (2.7\,km\,s$^{-1}$). The results of these models were compared to the spectra of the most optically thin lines to determine the best-fit model by computing the minimum $\chi^2$.\\

The brightness temperature of the line emission is calculated by CASSIS according to:
\begin{equation}
T_B(\nu) = T_0\left(\frac{1}{e^{T_0/T_{\rm ex}}-1} - \frac{1}{e^{T_0/T_{\rm bg}}-1} \right) (1 - e^{-\tau_{\nu}}),
\end{equation}
where $T_0 = h\nu/k$.
A derivation of this formula using the equation of radiative transfer can be found in \citet{Vastel2016}. Typically the background temperature ($T_{\rm bg}$) is the CMB background of 2.73\,K. Both sources in IRAS 16293, however, exhibit bright continuum emission from the dense dust in each source. This means a significant background radiation temperature should be considered when calculating the brightness temperature of the line emission. This can be applied to the column densities as a correction factor. The correction factor is 1.1 at 150\,K towards IRAS 16293B.\\

The initial modelling of the methyl isocyanide emission lines was performed using the JPL catalog \citep{Pickett1998} entry. Lines were assigned tentatively, but the predictions had large uncertainties (more than 0.5\,MHz or 0.4\,km\,s$^{-1}$), deviated to lower frequencies by almost one line width at $K = 0$, rapidly increasing to higher $K$. The JPL entry was based on \citet{MeNC_rot_1970} with an estimate of the rotational parameter $A-B$. These early laboratory data, however, extend only to 141\,GHz with $J = 7-6$ and $K \le 6$. A recent study provides additional rotational data up to 643\,GHz supplemented by ground state combination differences from the $\nu _4$ infrared (IR) spectrum \citep{MeNC_gs_v4_2011}. The rotational parameter $A-B$ and the centrifugal distortion parameter $D_K$, necessary for describing the $K$ level spacing, were taken from an earlier IR study \citep{MeNC_IR_A_1995}. The resulting predictions will be made available in the Cologne Database for Molecular Spectroscopy\footnote{\url{http://www.astro.uni-koeln.de/cdms}} (CDMS; \citealt{Endres2016}). Vibrational contributions to the partition function at 150\,K were estimated from $\nu _8$ and its overtones from that earlier IR study as 1.17.

An error analysis was also carried out for the column densities, excitation temperatures and abundances determined in this work. The flux calibration accuracy of these data are better than 5\% \citep{Jorgensen2016}. A larger source of error comes from the quality of the fit of the spectral model to the data. We therefore determine the errors by determining the minimum change to each quantity which produces a significant difference to the goodness of fit. \\

\begin{figure*} 
 \begin{center} 
  \includegraphics[width=18.5cm, angle=0, clip =true, trim = 2cm 5cm 2cm 2cm]{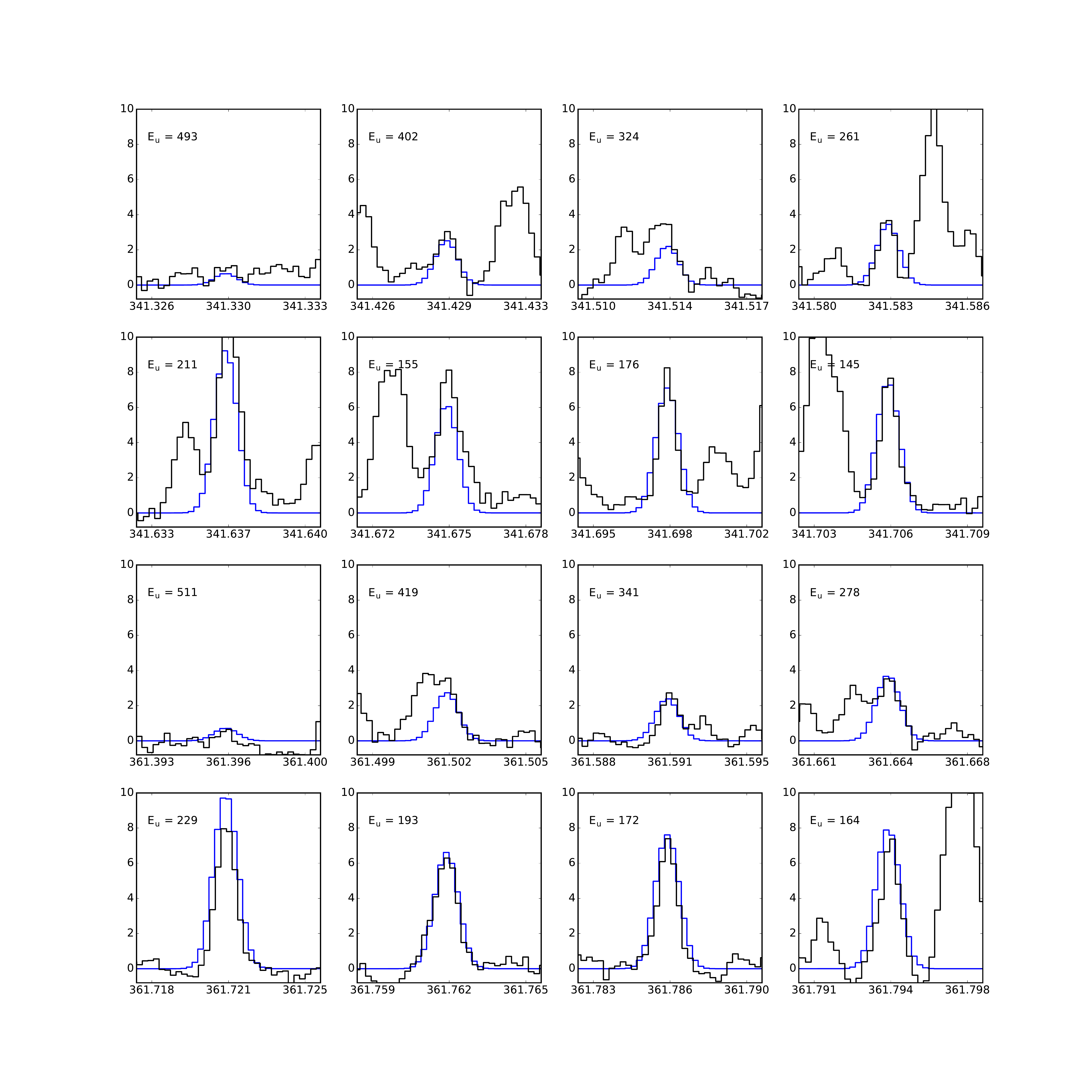}\\
{\fontfamily{ss}\selectfont \footnotesize{Frequency (GHz)}}
\begin{picture}(10,10)
{\fontfamily{ss}\selectfont \footnotesize{ \put(-260,200){\rotatebox{90}{Brightness temperature (K)}}}}

\end{picture}
 \end{center} 
  \vspace{-0.5cm}
 \caption{The lines of CH$_3$NC detected in IRAS 16293B (black) overlaid with an LTE spectral model (blue). The upper energy level of each line is given in the top left corner of each plot in K.\label{fig:ch3nc}} 
 \end{figure*}

Table \ref{tab:coldens} shows the best-fit excitation temperature and column density for methyl isocyanide in IRAS 16293B and the upper limits for IRAS 16293A, as well as the CH$_3$CN/CH$_3$NC abundance ratio using the methyl cyanide abundances determined in \citet{Calcutt2017}. Fig. \ref{fig:ch3nc} shows the LTE model overlaid on the data for all detected transitions in IRAS 16293B overlaid with the best-fit LTE model. The upper limits are determined for the brightest methyl isocyanide line in the frequency range of the observations (the 17$_3$\,--\,16$_3$ line at  341.637\,GHz). We use $1.05{\times}3{\times}RMS{\times}\sqrt{{\Delta}V{\times}FWHM}$ to compute the 3$\sigma$ limit, where 1.05 is a factor to account for a 5\% flux calibration uncertainty.  We assume a line width of 2.2\,km\,s$^{-1}$ based on the line width found for nitriles towards this source by \citet{Calcutt2017}, and an excitation temperature of 150\,K. The CH$_3$CN/CH$_3$NC abundance ratio in IRAS 16293B is 200 which is much lower than that found in IRAS 16293A (>5517).

\begin{table} 
\caption{Excitation temperature ($T_{\rm ex}$), column density ($N\rm{_{tot}}$) and the abundance ratio, N(CH$_3$CN)/N(CH$_3$NC), in IRAS 16293B and IRAS 16293A.}\label{tab:coldens}
\centering
\footnotesize
\begin{tabular}{cccc}
\hline
\hline
Source&$T\rm{_{ex}}$&$N\rm{_{tot}}$  &\underline{$N$(CH$_3$CN)$^{c}$}\\
&(K)&(cm$^{-2}$)&$N$(CH$_3$NC)\\
\hline
IRAS 16293B&150$\pm$20&2.0$\pm$0.2$\times$10$^{14}$&200 \\
IRAS 16293A&150$^a$&$<$1.45$\times$10$^{13}$$^{b}$&$>$5317\\
\hline
\end{tabular}
\tablefoot{All models assume LTE and a source size of 0\farcs5. The FWHM of IRAS 16293B lines is 1$\pm$0.2\,km\,s$^{-1}$ and the peak velocity is 2.7$\pm$0.2\,km\,s$^{-1}$, based on a fit of the data. $^{a}$The excitation for IRAS 16293A is adopted to be the same as IRAS 16293B and similar to other molecules detected towards this source. $^{b}$Determined for the 3$\sigma$ upper limit assuming an  FWHM of 2.2\,km\,s$^{-1}$, based on the FWHM of other molecules detected in this source in \citet{Calcutt2017}. $^{c}$Determined using CH$_{3}$CN column densities from \citet{Calcutt2017}.}
\end{table}

\subsection{Chemical modelling} \label{sec:chemmod}

In order to investigate the chemistry of CH$_3$NC, we use an updated version of the three-phase chemical kinetics model {\em MAGICKAL} \citep{Garrod2013}. This model has been updated with the back-diffusion correction of \citet{Willis2017}. The chemical network is based on that of \citet{Belloche2017}, using gas-phase, grain-surface and bulk ice chemistry, with formation and destruction mechanisms for CH$_3$NC added. This is, to the authors' knowledge, the first time that CH$_3$NC has been incorporated into an astrochemical kinetics model. This network is a preface to a more complete chemical treatment for isocyanides (Willis et al., in prep.). The two-stage physical model used is also similar to that of \citet{Belloche2017}, in which a cold collapse is followed by a static warm-up to 400\,K. The cold collapse in the model has an isothermal gas temperature
(10\,K), and the dust temperature cools from an initial value of 16\,K to a final value of 8\,K. The chemical model is a single-point model, thus it has a uniform density. Following \citet{Coutens2018}, we run two models, using two different final densities for the collapse phase: $n_{\mathrm{H}}=6\times10^{10}$ cm$^{-3}$, corresponding to the continuum peak of IRAS 16293B and $n_{\mathrm{H}}=1.6\times10^{7}$ cm$^{-3}$, corresponding to the density of the filament between IRAS 16293A and IRAS 16293B \citep{Jacobsen2018}. The subsequent warm-up phase starts at a dust temperature of 8 K, and reaches a final temperature of 400 K at $2.8\times10^{5}$ years.  This timescale is used to represent an intermediate-timescale warm-up, where 2$\times$10$^{5}$ years is the time spent to reach a dust temperature of 200\,K. This is taken from \citet{Garrod2006}. The model presented here follows \citet{Garrod2013} in using a warm-up to 400\,K instead of 200\,K, by extending the ``intermediate'' temperature function beyond 200\,K.

The primary formation mechanism of CH$_3$NC in our model follows \citet{Defrees1985}. First, the radiative association of CH$_3$$^+$ and HCN forms CH$_3$CNH$^+$. This intermediate is formed with enough energy that a certain percentage of these molecules can isomerise to CH$_3$NCH$^+$; we use the suggestion by DeFrees et al. of 15\%. CH$_3$NCH$^+$ then recombines with free electrons, forming CH$_3$NC and H as the primary channel ($\sim$65\% of recombinations), with the remainder of CH$_3$NCH$^+$ molecules recombining to HCN and CH$_3$. The binding energies on water ice for methyl cyanide and methyl isocyanide were updated in the network using new values from Bertin et al. (2017a; 6150 K and 5686 K, respectively). The primary destruction mechanism for CH$_3$NC on the grain surface is reaction with H, forming HCN and the radical CH$_3$. This reaction is assumed to have a barrier of 1200 K, based on recent calculations of the barrier for H + HNC \citep{Graninger2014}. The same reaction is also allowed to occur in the gas phase, assuming the same barrier.

Since we are comparing CH$_3$NC abundances to values for CH$_3$CN, it is important to review the how CH$_3$CN is formed in our network as well. CH$_3$CN is formed primarily via the hydrogenation of the CH$_2$CN radical on the grain surface and in the ice mantle, with a small contribution from reaction of CH$_2$CN with HCO. There is also a viable, less efficient gas-phase pathway to production of CH$_3$CN, previously discussed in regards to CH$_3$NC. The CH$_3$CNH$^+$ molecules that do not isomerise to CH$_3$NCH$^+$ can recombine with electrons to form CH$_3$CN and H as the primary products, with some instances of this channel leading to HNC and CH$_3$.

The basic chemical network we construct here for CH$_3$NC chemistry is intended to incorporate the most obvious formation and destruction mechanisms, in keeping with the larger network. It is nevertheless possible that alternative mechanisms exist, while the efficiencies of those included here cannot be reliably known without careful experimental studies. In particular, the reaction of H+CH$_3$NC, which is included in both the gas-phase and the grain-surface networks, is quite speculative. The mechanics of the reaction with H and HNC are fundamentally different from this process, but the lack of experimental work makes it difficult to produce a better prediction. In particular, there is a large uncertainty on the barrier of the CH$_3$NC+H reaction as it likely does not equal the barrier on the HNC+H reaction. We hope to continue to refine this network for CH$_3$NC and other isocyanides (Willis et al., in prep.), but this result highlights the need for more experimental work on these molecules.
\begin{figure*} 
\begin{center} 
\includegraphics[width=17cm, angle=0, clip =true, trim = 3.7cm 4.3cm 4cm 0cm]{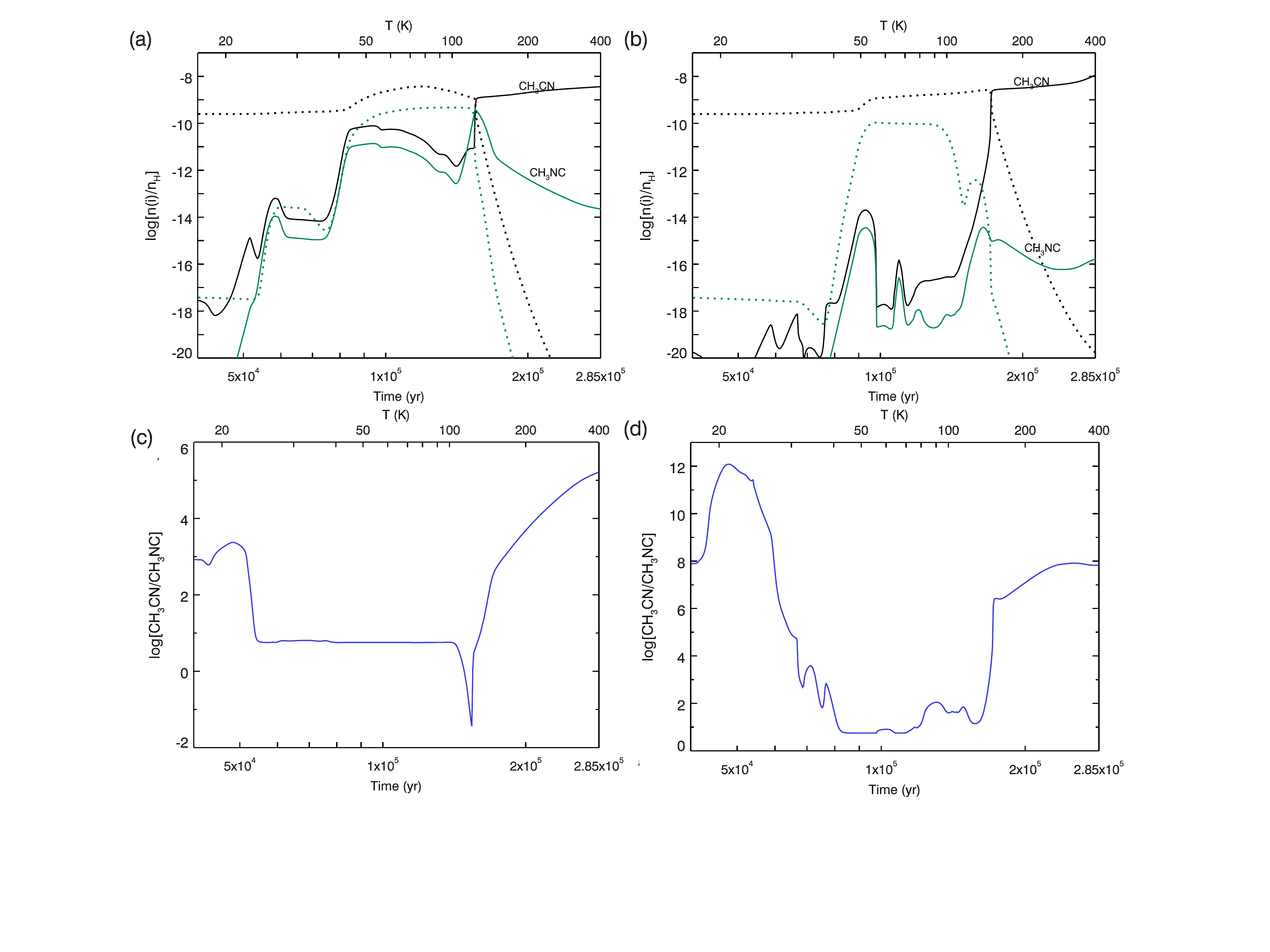}
\end{center} 
\caption{Chemical model abundances of CH$_3$CN (black) and CH$_3$NC (green) for the warm-up stage of a hot-core type model. Panel (a) is for a model with a final collapse density of $n_{\mathrm{H}}=1.6\times10^{7}$ cm$^{-3}$, and panel (b) is for a model with a final collapse density of $n_{\mathrm{H}}=6\times10^{10}$ cm$^{-3}$. Solid lines denote gas-phase abundances, while dashed lines indicate grain-surface abundances. Panel (c) shows the CH$_3$CN/CH$_3$NC ratio (blue) for the $n_{\mathrm{H}}=1.6\times10^{7}$ cm$^{-3}$ model, and panel (d) shows the CH$_3$CN/CH$_3$NC ratio (blue) for the $n_{\mathrm{H}}=6\times10^{10}$ cm$^{-3}$ model.
 \label{fig:ch3ncmodel}} 
\end{figure*} 
The results of the models are shown in Fig. \ref{fig:ch3ncmodel}, with panel (a) corresponding to $n_{\mathrm{H}}=1.6\times10^{7}$ cm$^{-3}$, panel (b) to $n_{\mathrm{H}}=6\times10^{10}$ cm$^{-3}$, and panels (c) and (d) showing the CH$_3$CN/CH$_3$NC ratio for each of the models. The different densities exhibit different behavior, particularly with regards to CH$_3$NC. The peak abundance of CH$_3$NC in the low-density model is $\sim$4$\times$10$^{-10}$, while the high-density model produces a much lower peak abundance, $\sim$4$\times$10$^{-15}$. This is due to the increased efficiency of grain-surface destruction of CH$_3$NC via reaction with H in the high-density model. As a result of this, as well as a higher peak abundance of CH$_3$CN in the high-density model, the CH$_3$CN/CH$_3$NC ratios at 150 K (the approximate excitation temperature of CH$_3$NC) vary greatly between the two models. In the low-density case, the ratio at 150 K is $\sim$450, which is the same order as the observed ratio in IRAS 16293B. The ratio at the same point in the high-density model is $\sim$2.5$\times$10$^{6}$, which is consistent with the upper limits for IRAS 16293A. 

We also ran several models incorporating the radiative association of CH$_{3}$$^{+}$ and HNC to test the effect that this proposed reaction has on the abundances of CH$_3$CN and CH$_3$NC. This process was given the same rate as the equivalent association of HCN with CH$_{3}$$^{+}$, and results in CH$_{3}$NCH$^{+}$. This ion is then assumed to have enough internal energy to isomerise as in the HCN reaction. Several different branching ratios were tested, including those that lead predominantly to CH$_{3}$CN and CH$_{3}$NC. However, it was observed that the inclusion of this process did not affect the abundance of CH$_{3}$CN in any significant way, and had only a small effect on the abundance of CH$_{3}$NC in the low-density model, increasing the peak abundance to a maximum of $\sim$$9\times10^{-10}$ when the reaction favours CH$_3$NC production.

\begin{figure} 
\begin{center} 
\includegraphics[width=8cm, angle=0, clip =true, trim = 10cm 0cm 10cm 0cm]{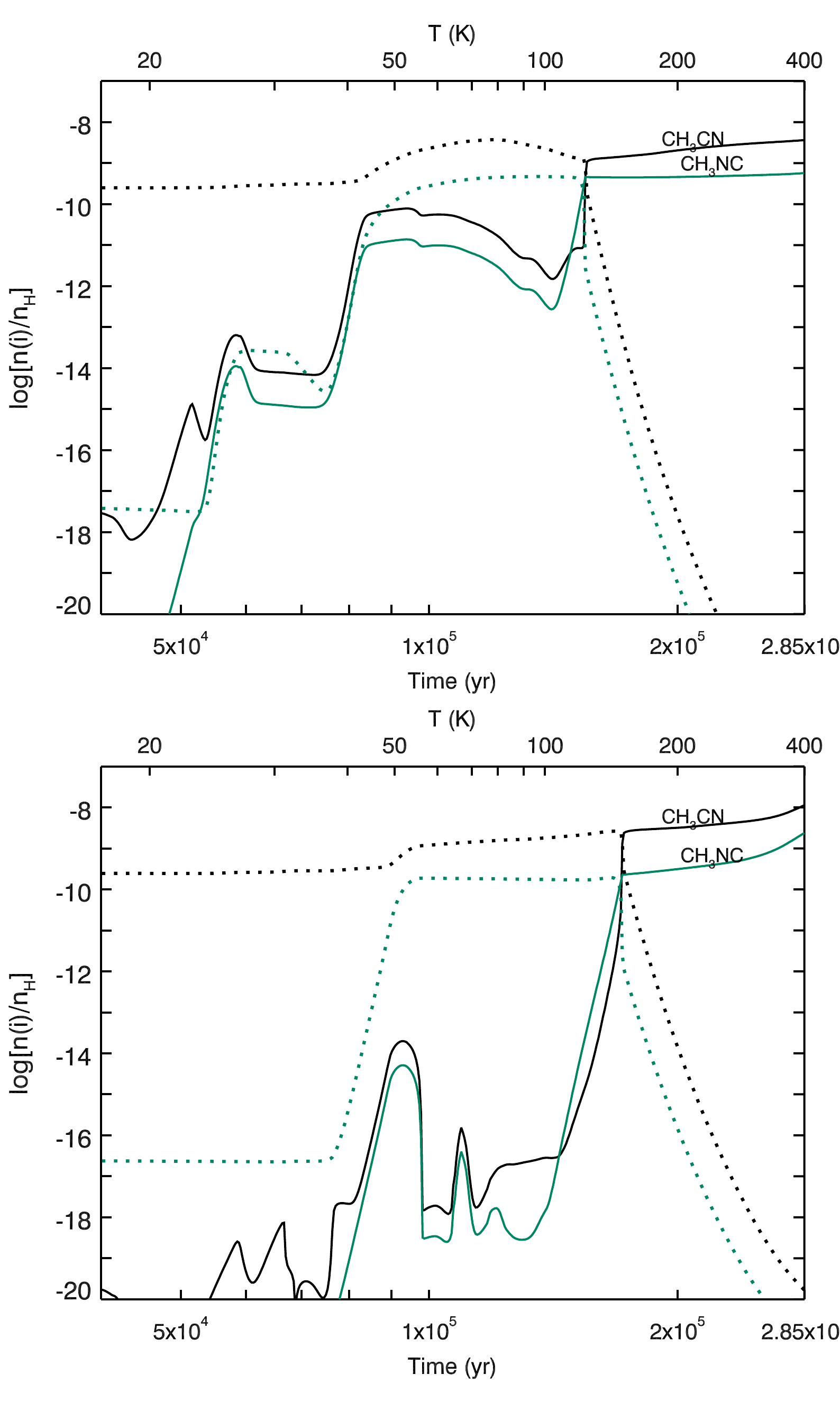}
\end{center} 
\caption{Chemical model abundances of CH$_3$CN (black) and CH$_3$NC (green) for the warm-up stage of a hot-core type model where the main destruction pathway for CH$_3$NC (H+CH$_3$NC) has been disabled. Solid lines denote gas-phase abundances, while dashed lines indicate grain-surface abundances. The top panel shows the model with a final collapse density of $n_{\mathrm{H}}=1.6\times10^{7}$ cm$^{-3}$ and the bottom panel shows the model with a final collapse density of $n_{\mathrm{H}}=6\times10^{10}$ cm$^{-3}$.
 \label{fig:ch3nch}} 
\end{figure}

Additionally, we have explored disabling the H+CH$_3$NC grain surface reaction and gas-phase reaction, to understand how important this speculative reaction is on the CH$_3$CN/CH$_3$NC ratio. Disabling the grain surface reaction had practically no effect, however, disabling the gas-phase reaction had a significant impact at higher temperatures (Fig. \ref{fig:ch3nch}). CH$_3$NC is not destroyed efficiently in the gas-phase and this keeps the CH$_3$CN/CH$_3$NC ratio much closer to unity. It is in fact quite close to the 85\%/15\% split seen in the formation pathways.  This result shows that there needs to be some efficient mechanism for destruction of CH$_3$NC in the gas-phase that doesn't operate on CH$_3$CN. 
The CH$_3$CN/CH$_3$NC ratio varies greatly in the temperature range of 130\,$-$\,170\,K. This is due to multiple factors. First, both species come off of the grain surface in this window, with CH$_3$CN coming off slightly later due to its somewhat higher binding energy. Once both species are off the grain, CH$_3$NC can be destroyed in the gas phase by reaction with H, which is not the case for CH$_3$CN. This leads to a dip in CH$_3$NC abundance relative to CH$_3$CN and a change in the abundance ratio of these molecules. This may be the reason for the difference in CH$_3$CN/CH$_3$NC between sources A and B. Also, although the low-density model fits the observations better than the high-density model, it may not be a representative density for the regions being probed by these observations. This and recent results from our other models \citep{Coutens2018} highlight the need for chemical models in which the collapse and warm-up phases occur simultaneously, and which are tailored for specific objects. Models in which the maximum density is reached prior to any warm-up seem to provide a poor representation of the chemistry in cases where densities reach extreme values. This density threshold appears to be around $10^{9}$ cm$^{-3}$, based on results from Coutens et al. for NH$_2$CN. 

 \section{Discussion}\label{sec:dis}
The lack of detection of methyl isocyanide in the second component in the IRAS 16293 binary, IRAS 16293A, shows that there is significant chemical differentiation between these sources. It is difficult to determine how common such differences are for different protostars, with only a handful of methyl isocyanide detections in the literature available for comparison. Table \ref{tab:othersources} shows the CH$_3$CN/CH$_3$NC abundance ratio from these detections in low-mass hot corinos, high-mass hot cores, intermediate-mass hot cores, cold cores, and PDRs. Several of these ratios are based on either tentative detections or on abundance values that are not robust, making a detailed comparison with them problematic, but they can give a general picture of the chemistry. The CH$_3$CN/CH$_3$NC abundance ratio for IRAS 16293B falls in the middle of the range of values determined in intermediate and high-mass sources, and is an order of magnitude higher than in the TMC-1 cold core and the Horsehead PDR. The value in IRAS 16293A is more than an order of magnitude higher than in the other hot core/hot corino objects.

Comparisons of the spatial scale of methyl isocyanide to other hot core/hot corino objects is limited due to a lack of information in this regard. For the most part it is assumed to be co-spatial with methyl cyanide emission, emanating from the hot core component as is found in IRAS 16293B. A different scale is found in Sgr B2(N) by \citet{Remijan2005}, where it is deficient in the hot core component, showing only large-scale emission. Such differences between Galactic Centre objects and objects outside the Galactic Centre is unsurprising with the significant differences in the physical environments of such objects. In fact, differences in the spatial scale between Galactic Centre objects and objects outside the Galactic Centre are also seen for other molecules such as glycolaldehyde \citep{Beltran2009, Hollis2004}.
 
The chemical modelling presented in Section \ref{sec:chemmod} fails to reproduce the observed CH$_3$CN/CH$_3$NC abundance ratio at the high densities of IRAS 16293 B, but the abundance ratio produced is consistent with the upper limits of IRAS 16293A. It does show that the ratio depends on the temperature, and consequently evolutionary stage of the source. An earlier stage of evolution would correspond to a lower CH$_3$CN/CH$_3$NC ratio, when CH$_3$NC has not been significantly destroyed by the gas-phase reaction with H. This combined with additional formation mechanisms which have not been included in the model at present could result in abundance ratios closer to the observational values. Such a scenario, if applied to IRAS 16293, suggests that the B source is younger than the A source or at least less evolved. This agrees broadly with the outflow picture of the two sources, where outflows have only been detected from IRAS 16293A \citep{Kristensen2013, Girart2014}, and not IRAS 16293B, where it may not have had time to establish outflows. It is, however, difficult to disentangle the large scale emission around the B source and therefore determine if there are any indications of previous outflow activity from the source. It could be that B is the less evolved source and therefore its outflows have not yet `switched on' or that it is currently going through a quiescent accretion/ejection phase.

This view of the evolutionary stage of these two sources is contradicted by the vinyl cyanide C$_2$H$_3$CN abundances towards both sources. \citet{Garrod2017} find that the vinyl cyanide abundance only becomes significant during the late stages of the warm-up phase of a protostar's formation. \citet{Calcutt2017} find vinyl cyanide to be 9 times more abundant towards IRAS 16293B, which could indicate that the B source is the more evolved protostar. They also suggest that both sources have the same physical age and mass but that the A source has higher accretion rates, leading to a warm-up timescale so short that vinyl cyanide is not efficiently formed. At present it is difficult to determine whether either scenario is describing the IRAS 16293 system. 

\section{Conclusions}\label{sec:con}
In this work we have presented the first detection of methyl isocyanide in a solar-type source, IRAS 16293B, and the first chemical modelling of methyl isocyanide. This represents a significant step forward in understanding methyl isocyanide chemistry in the ISM. Many key questions, however, still remain in order to explain the observed CH$_3$CN/CH$_3$NC abundance ratios in different objects. In particular, is the isomerisation of CH$_3$CNH$^+$ the dominant reaction pathway to form this molecule? The chemical modelling of methyl isocyanide has also shown that the gas-phase destruction mechanism, H+CH$_3$NC, is necessary to increase the CH$_3$CN/CH$_3$NC ratio to those found in hot corino/core objects. Whilst this reaction is only speculative, having not yet been studied experimentally and has a very uncertain barrier, it does highlight the need for an efficient mechanism for the destruction of methyl isocyanide in the gas-phase that doesn't operate on methyl cyanide.

The chemical modelling has also shown that the CH$_3$CN/CH$_3$NC abundance ratio is very sensitive to the final collapse density. At the high densities representative of both A and B, this results in higher CH$_3$CN/CH$_3$NC abundance ratios which do not match the observed ratio towards source B. Only the upper limits towards the A source are consistent with the chemical modelling in the high density case.

The next step to understanding methyl isocyanide chemistry is further detections of methyl isocyanide, to establish its abundance ratio in a number of objects. Such detections will also need to be compared to further experimental and theoretical work, to explore other formation and destruction mechanisms, and build a more complete picture of this molecule in the ISM. Crucially, to explore methyl isocyanide chemistry with chemical modelling a more comprehensive physical model is needed, where warm-up is performed after the final density has been reached. This is especially crucial in objects such as IRAS 16293 where densities reach extreme values.

 \begin{table}

	\caption{CH$_3$CN/CH$_3$NC ratios in different objects}\label{tab:othersources}
	\centering
	\footnotesize
	\begin{tabular}{cccc}
		\hline
		\hline
		Source & \underline{$N$(CH$_3$CN)} &Object type& Ref.\\
		&$N$(CH$_3$NC)&&\\
		\hline
		IRAS 16293A & $>$5517$^{\dagger}$ &LM& -\\
		IRAS 16293B & 200 & LM&-\\
		Sgr B2(N) molecular cloud & 50&HM &[1]\\
		Orion KL & 500 & HM&[2]\\
		W51 e1/e2 & 96$^{\star}$ &HM&[3]\\
	        NGC 7129 FIRS 2 & 350$^{\star}$&IM &[4]  \\
		TMC-1 & \begin{math}\geq \end{math}  11 &CC& [5] \\
		Horsehead& 7&PDR&[6]\\
		\hline
		
	\end{tabular}
	\tablefoot{
	$^{\dagger}$Based on upper limits of CH$_{3}$NC. $^{\star}$CH$_{3}$NC only tentatively detected. LM stands for low-mass hot core, HM stands for high-mass hot core, IM stands for intermediate-mass, CC stands for cold core, and PDR stands for photon-dominated region.}
	\tablebib{
(1) \cite{Remijan2005}; (2) \cite{Lopez2014}; (3) \cite{Kalenskii2010}; (4) \cite{Fuente2014}; (5) \cite{Irvine1984}; (6) \cite{Gratier2013}}
\end{table}

\begin{acknowledgements}
The authors would like to thank the referee for their excellent suggestions on this work. They would also like to thank Ewine van Dishoeck for her helpful comments and suggestions. The authors would like to acknowledge the European Union whose support has been essential to this research. In particular a European Research Council (ERC) grant, under the Horizon 2020 research and innovation programme (grant agreement No. 646908) through ERC Consolidator Grant "S4F" to J.K.J. Research at the Centre for Star and Planet Formation is funded by the Danish National Research Foundation. Astrochemistry in Leiden is supported by the European Union A-ERC grant 291141 CHEMPLAN, by the Netherlands Research School for Astronomy (NOVA) and by a Royal Netherlands Academy of Arts and Sciences (KNAW) professor prize. A.C. postdoctoral grant is funded by the ERC Starting Grant 3DICE (grant agreement 336474). MND acknowledges the financial support of the Center for Space and Habitability (CSH) Fellowship and the IAU Gruber Foundation Fellowship. This paper makes use of the following ALMA data: ADS/JAO.ALMA\#2013.1.00278.S. ALMA is a partnership of ESO (representing its member states), NSF (USA) and NINS (Japan), together with NRC (Canada) and NSC and ASIAA (Taiwan), in cooperation with the Republic of Chile. The Joint ALMA Observatory is operated by ESO, AUI/NRAO and NAOJ.
\end{acknowledgements}

\bibliographystyle{aa}
\bibliography{./ch3nc}

\end{document}